\documentclass{osa-article}

\journal{oe}

\articletype{Research Article}

\usepackage{lineno}
\usepackage{bm}

\begin{document}


\title{High-bandwidth CMOS-voltage-level electro-optic modulation of 780 nm light in thin-film lithium niobate}

\author{Oguz Tolga Celik,\authormark{1} Christopher J. Sarabalis,\authormark{2} Felix M. Mayor,\authormark{1} Hubert S. Stokowski, \authormark{1} Jason F. Herrmann, \authormark{1} Timothy P. McKenna, \authormark{1,3} Nathan R. A. Lee, \authormark{1} Wentao Jiang, \authormark{1} Kevin K. S. Multani, \authormark{1} and Amir H. Safavi-Naeini\authormark{1,*}}

\address{\authormark{1}Ginzton Laboratory, Stanford University, 348 Via Pueblo Mall, CA 94305, USA\\\authormark{2} Flux Photonics Inc., 580 Crespi Dr Unit R Pacifica, CA 94044 USA\\\authormark{3} NTT Research, Inc., 940 Stewart Dr Sunnyvale, CA 94085 USA\\
}

\email{\authormark{*}safavi@stanford.edu} 



\begin{abstract}
Integrated photonics operating at visible-near-infrared (VNIR) wavelengths offer scalable platforms for advancing optical systems for addressing atomic clocks, sensors, and quantum computers. The complexity of free-space control optics causes limited addressability of atoms and ions, and this remains an impediment on scalability and cost. Networks of Mach-Zehnder interferometers can overcome challenges in addressing atoms by providing high-bandwidth electro-optic control of multiple output beams. Here, we demonstrate a VNIR Mach-Zehnder interferometer on lithium niobate on sapphire with a CMOS voltage-level compatible full-swing voltage of 4.2 V and an electro-optic bandwidth of 2.7 GHz occupying only 0.35 mm$^2$. Our waveguides exhibit 1.6 dB/cm propagation loss and our microring resonators have intrinsic quality factors of $4.4\times 10^5$. This specialized platform for VNIR integrated photonics can open new avenues for addressing large arrays of qubits with high precision and negligible cross-talk.
\end{abstract}

\section{Introduction}

Scaling atom and ion quantum computers requires scalable control optics. Multi-channel acousto-optic modulators have been used for the addressed control of tens of ions \cite{BenchmarkIon, FastIon}, but scaling to hundreds remains as a challenge.  Similarly, hundreds of neutral atoms can be trapped, cooled, and measured in optical tweezer arrays \cite{microOpticalArray, atomicEnsembles, atomAssembly}, but addressed control of these atoms is slow and serial. Scaling to the addressed control of hundreds of atoms and ions or beyond calls for a new architecture for the control optics.

Integrated photonics platforms operating at visible-near-infrared (VNIR) wavelengths possess great potential for applications in a wide variety of fields in atomic, molecular and optical (AMO) physics. In particular, alkali and alkaline-earth metals, such as rubidium, cesium, calcium and strontium, which are essential for trapped-ion \cite{trappedIons, highFidelityIon, RepeatedMultiqubit} and neutral atom \cite{QuantumMemory, Madjarov2020, HighFidelityRydberg} quantum computing have transitions in the VNIR range. Bulk VNIR light modulators have been used extensively for the control of trapped ions \cite{BenchmarkIon, FastIon} and Rydberg atoms \cite{GenerationRydberg, AnalysisImperfection, RydbergLevine} with the addressing schemes being limited to individual qubits. Integrated photonics can overcome current limitations imposed by control optics to achieve simultaneous control of multiple ions or atoms and enable a scalable framework for AMO quantum computation \cite{IntegratedMultiIon, IntegratedOpticsIon, MaterialsIon}.

Building a scalable platform to realize this vision requires low-loss, high-bandwidth control of light at visible-near-IR wavelengths. Electro-optic modulators allow high bandwidth control at telecom wavelengths, but examples operating at VNIR wavelengths are limited. Scalability of electro-optics depends on achieving low switching voltages (< 5 V), small device footprints ($<1$ mm$^2$), radio frequency (RF) and optical power handling ($>1$ mW), compatibility with dense wiring, and control over the optical mode-mixing for maximal on-off ratios. 

Developing visible integrated photonics has proven challenging as it requires moving beyond well-established CMOS fabrication techniques available to silicon photonics, since silicon is not transparent at the wavelengths of interest. Several material platforms for visible and near-IR integrated photonics such as  SiN \cite{SandiaSiN, SiNThermoOptic, SiNReview, SiN_CMOS}, AlGaN/AlN \cite{AlGaN_AlN, AlN_visible}, diamond \cite{DiamondNonlinear, DiamondPIC, DiamondNano}, SiO$_2$ \cite{sio2visible, sio2soliton}, GaP \cite{GaP_photonics} and lithium niobate \cite{Loncar_visible, li2022high} have been investigated. Among these, scalable piezo-optomechanical phase modulation was demonstrated using SiN and AlN at 100 MHz \cite{SandiaSiN} with a high voltage-length product of $V_{\pi}L \sim 50-60$ V-cm, but neither of these platforms, excluding the weaker ($\sim$ 1 pm/V) electro-optic effect of AlN \cite{AlN_vpiL}, are capable of efficient and high bandwidth electro-optic switching.

Lithium niobate (LN) is a particularly attractive material due to its wide transparency window (350 - 5000 nm), strong $\chi^{(2)}$ nonlinearity, electro-optic, piezo-electric and acousto-optic properties \cite{Weis1985} that make integration of various physical devices on one chip possible. Electro-optic intensity modulators in thin film lithium niobate (TFLN) are proven to be capable of achieving low-loss optical transmission \cite{monolithicQ, LN1064EO, Mercante:16}, CMOS-level switching voltages \cite{Wang2018} and ultra-high bandwiths (> 100 GHz) \cite{100GhzLN} at telecom and infrared (IR) wavelengths. In particular, these results were achieved using the LN-on-insulator (LNOI) platform where TFLN is bonded on a silicon or LN wafer with a silicon dioxide buffer layer \cite{LNOI_status}.

We demonstrate a VNIR integrated photonics platform on TFLN that satisfies all of the aforementioned requirements for scalable electro-optic modulator networks. High bandwidth integrated visible wavelength modulation has been previously demonstrated by Desiatov et al. \cite{Loncar_visible} on the LNOI platform. Our work pays particular attention to improved performance in anticipation of larger scale systems. In particular, we focus on lower modulation voltages, improved RF power handling, high extinction, and realization of lower-loss optical components that can be cascaded without significant detrimental loss, scattering, or cross-talk.

An alternative to LN-on-SiO$_2$ (LNOI) is the LN-on-sapphire (LiSa) platform, which exhibits wide transparency up to Mid-IR \cite{MidIRLiSa}, supports high Q-factor microwave resonance \cite{SapphireMicrowave} and LN-guided acoustic waves that enable on-chip acousto-optics \cite{chris} and phononics \cite{phononics_lisa}. Sapphire also has high thermal conductivity, which translates to better RF power handling capability, and has lower RF loss than SiO$_2$, which is important for cryogenic applications, such as microwave-to-optical conversion \cite{cryogenic}. LN ($n\approx 2.2$) and sapphire ($n\approx 1.7$) have an index contrast that is, although smaller than that of LNOI, sufficient to enable high confinement in waveguides, and realize localization to make efficient electro-optic modulation possible. In addition to material selection, it is critical to have minimal optical reflection and mode mixing to achieve high output contrast with deep photonic circuits. To minimize reflections we use directional couplers as beamsplitters, and to minimize mode mixing we conduct a careful analysis of the band structure. We fabricate through-oxide vias (trenches) to allow dense wiring and reduce photorefractive effects. We demonstrate a low-loss integrated optics platform for a near-IR wavelength of 780 nm (tuned at the rubidium $5S_{1/2} \rightarrow 5P_{3/2}$ transition) on LiSa and use this platform to realize intensity modulators with high EO bandwidth (2.7 GHz) and CMOS-compatible switching voltages (4.2 V).

\section{Lithium Niobate on Sapphire Visible Integrated Optics Platform}

\subsection{Device Overview}
\begin{figure}[!ht]
\centering\includegraphics[width=0.7\textwidth]{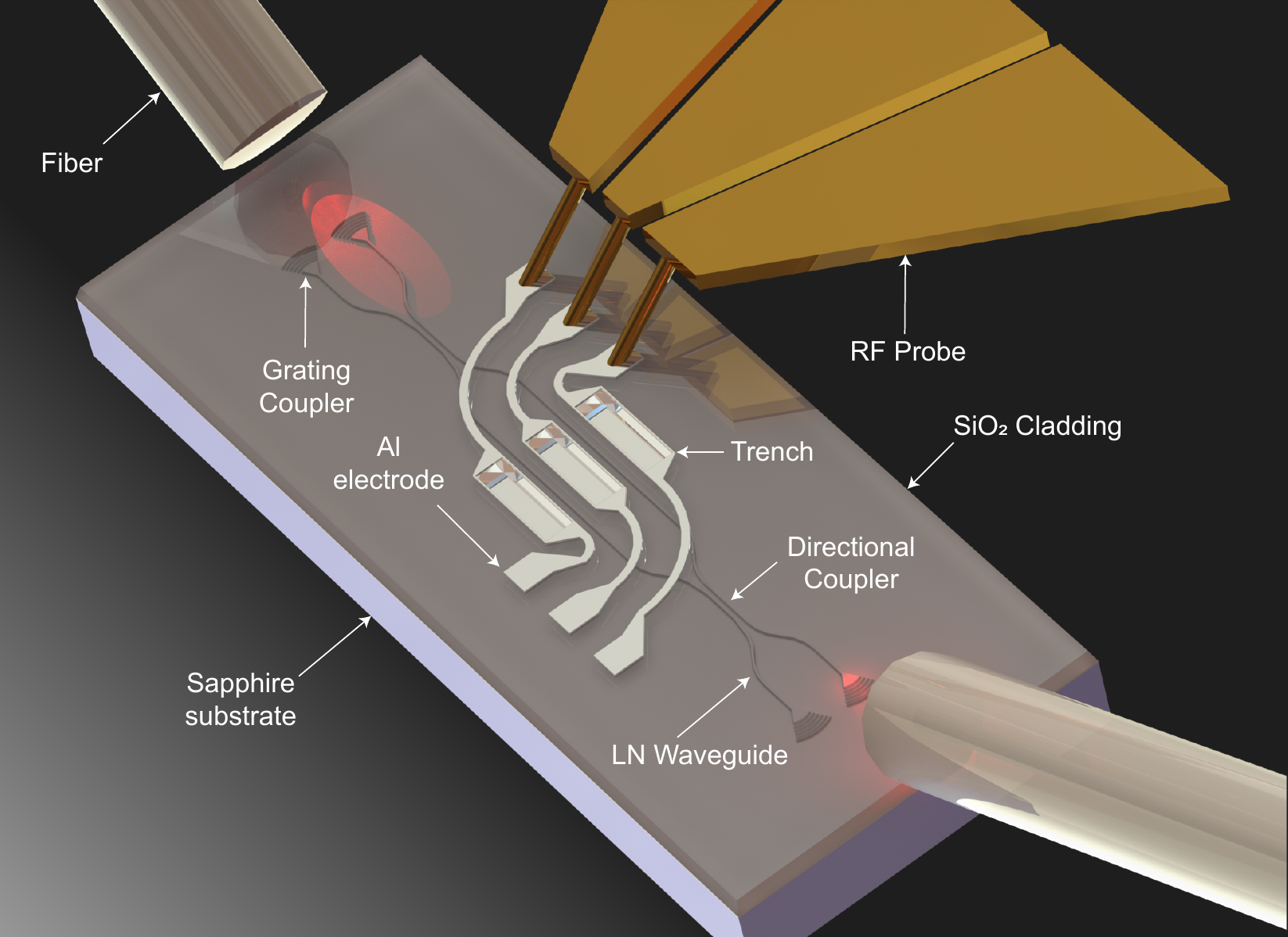}
\caption{Device layout and measurement scheme. VNIR light couples into the waveguides through grating couplers, and the directional coupler splits the beam 50:50 into the MZI arms. The GSG RF probe delivering the modulation signal is positioned perpendicular to the fibers on the measurement setup, so the Al electrodes make crossovers above the waveguides that lead to trenches in the oxide cladding, where the electrodes make contact with the LN. The second directional coupler combines the beams from the two arms of the MZI, and a grating couples the light to the output fiber by focusing on-chip light into a collimated Gaussian beam.}
\label{fig1:full-stack}
\end{figure}

In this section we introduce the visible light modulator in the form of an electro-optically tunable Mach-Zehnder Interferometer (MZI). Fig. \ref{fig1:full-stack} shows the overview of the MZI, which consists of beam splitters and electro-optic phase modulators \cite{SAGHAEI2019100733}. We implement 50:50 beam splitters as directional couplers centered at a wavelength of $\lambda = 780$ nm with a tapered coupling region to reduce the device footprint. The directional couplers are designed to have large bandwidth ($50\pm 5\%$ splitting ratio within 765-787 nm, corresponding to an extinction ratio $>20$ dB  on the MZI). Moreover, they minimize reflections and ensure low background on transmission. Aluminum (Al) electrodes deliver the modulation signal from a GSG microwave probe to both MZI arms that serve as phase modulators situated in a push-pull configuration. In order to prevent optical absorption, we distance the Al electrodes from the waveguides by patterning crossovers above the oxide cladding. We use 10\% efficient grating couplers for optical characterization and couple light into and out of the chip through two flat-cleaved fibers. Photorefractive (PR) effects limit the power handling of modulators by shifting the effective mode index and limiting the EO modulation through electric field screening \cite{Weis1985}; therefore it is crucial to limit PR effects by providing a migration path to accumulating static charge or removing the cladding layer to eliminate interfacial trap sites \cite{photorefractive_LN}. To reduce photorefractive effects, we etch trenches in the cladding and bring the electrodes in contact with the LN layer.

\subsection{Optical Design and Fabrication} 
\begin{figure}[!ht]
\centering\includegraphics[width=\textwidth]{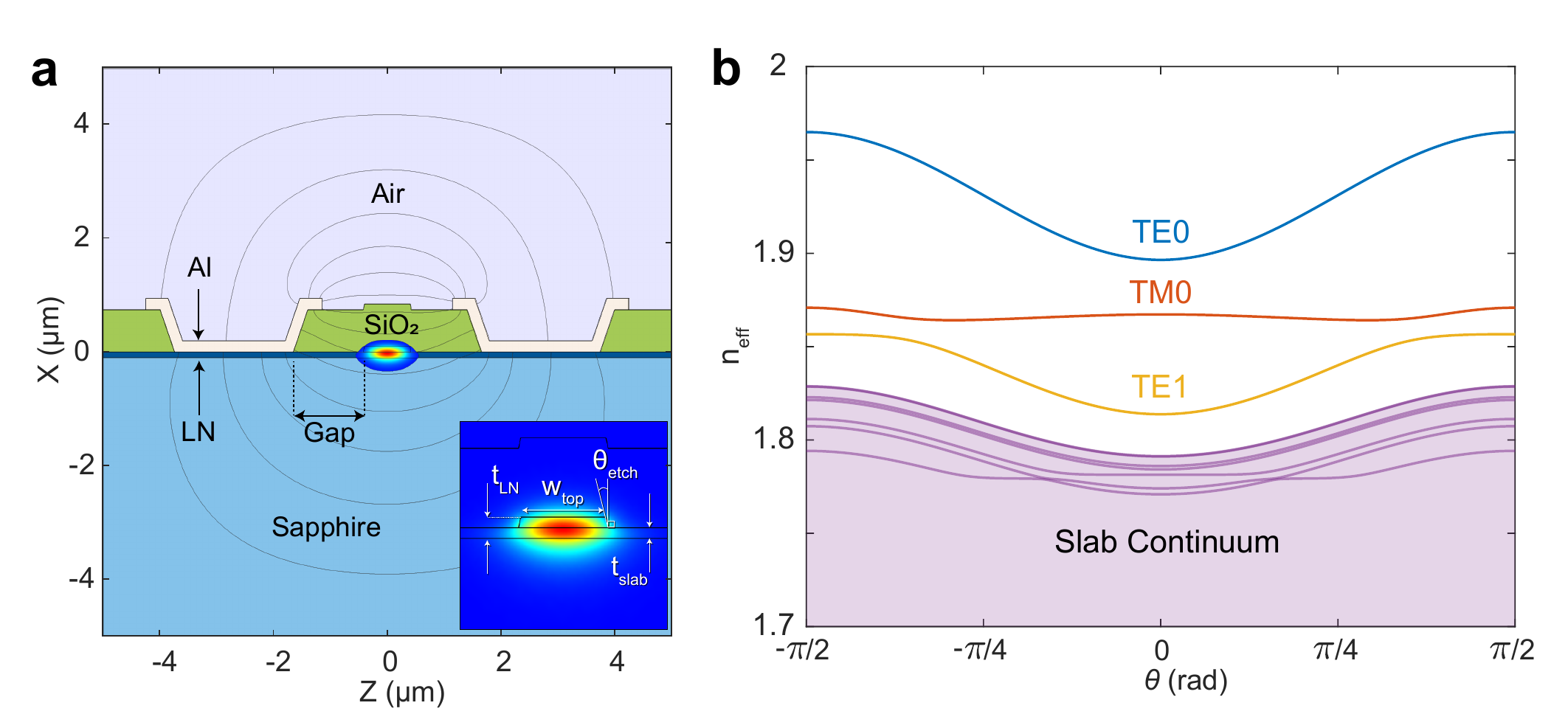}
\caption{\textbf{a.} Material stack of the electro-optic modulator along with $\text{TE}_0$ modal power density and E-field lines of the modulation field. Inset: $\text{TE}_0$ Mode profile and waveguide geometry design parameters. $\text{t}_{\text{LN}}$ is the total lithium niobate thickness, $\text{t}_{\text{slab}}$ is the slab thickness, $\text{w}_{\text{top}}$ is the top width of the waveguide and $\theta_{\text{etch}}$ is the sidewall angle. \textbf{b.} Effective index $\text{n}_\text{eff}$ with respect to propagation direction ($\theta$ is the positive angle with respect to the +Y axis in the YZ-plane).}
\label{fig2:optics-design}
\end{figure}

To realize deep, complex on-chip optical networks, it is crucial to build a low loss optical platform. Therefore, our design considerations focus on mitigating two main limitations related to the control of on-chip light: (i) propagation losses and (ii) intermodal scattering. We use ridge waveguides on X-cut LN with a silica cladding. To achieve low propagation losses, the following criteria guide our waveguide geometry design at the desired wavelength, $\lambda = 780$ nm: (i) single-mode operation for transverse electric (TE) and transverse magnetic (TM) modes, (ii) high mode confinement to reduce scattering from the waveguide sidewall and metal absorption, and (iii) minimal radiation losses due to bending. We design the optical waveguide geometry and the device stack as given in Fig. \ref{fig2:optics-design}.a to satisfy these conditions using mode analysis in COMSOL as shown in the inset.

Scattering between different modes of the waveguide occurs due to changes in the refractive index profile along the propagation length. A significant source of such intermodal scattering is from spatially-dependent inhomogeneity that occurs in a bend. As the crystal angle of a birefringent medium such as LN rotates, the local overlap of different modes causes them to mix, resulting in a significantly degraded performance of modulator networks by generating a non-modulated background and complex interference phenomena. Cascading transmission matrices shows that for a wave launched in the $\text{TE}_0$ mode with only 0.1\% intermodal power scattering between $\text{TE}_0$ and $\text{TM}_0$ modes per bend, results in 35\% of the total power propagating in the $\text{TM}_0$ mode after 20 bends. Since the modulating E-field along the Z-axis couples to $\text{TE}_0$ through the $\text{r}_{33}$ component of the electro-optic tensor and negligibly affects $\text{TM}_0$ in this configuration, the non-modulated $\text{TM}_0$ would result in a background that reduces the extinction ratio of the MZI. To mitigate this effect we design the waveguide geometry with careful consideration of the angle-dependent variation in the effective indices. We then use Local Coupled Mode Theory (LCMT) calculations, with COMSOL used for the mode profiles and overlap integrals, to more quantitatively design and evaluate the bends \cite{snyder2012optical}. We calculate the effective index as a function of crystal orientation by sweeping the propagation direction at 780 nm as shown in Fig. \ref{fig2:optics-design}.b. We choose the geometry such that the $\text{TE}_0$ has an effective index larger than the $\text{TM}_0$ mode for all values of the crystal angle. This eliminates any anticrossing between the two bands so that the intermodal scattering can be suppressed with sufficiently slow variation of the crystal angle $\theta$. We tolerate the emerging $\text{TE}_1$ mode to increase mode confinement and validate that there is no power scattering into this mode through LCMT studies. $\text{TE}_1$ power is concentrated close to the waveguide sidewalls and is therefore highly lossy, so the initial power coupled to this mode presents negligible background at the output. We find that given an etch angle constraint $\theta_{\text{etch}}  = 12^{\circ}$, the optimal design point for the waveguide geometry shown in the inset of Fig. \ref{fig2:optics-design}.a is $\text{w}_{\text{top}}  = 800$ nm, $\text{t}_{\text{LN}}  = 200$ nm, and $\text{t}_{\text{slab}} = 100$ nm.  

\begin{figure}[!ht]
\centering\includegraphics[width=\textwidth]{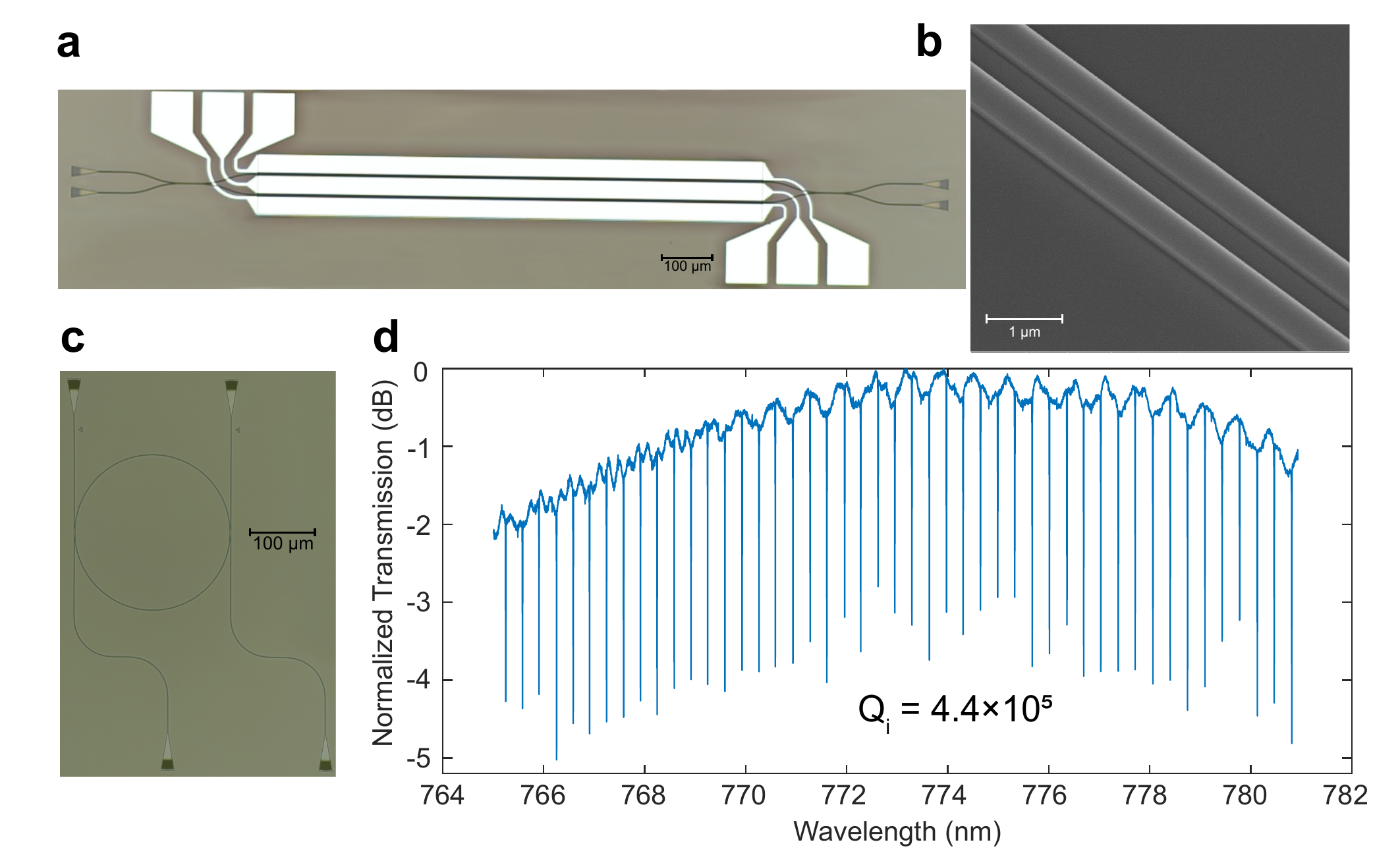}
\caption{\textbf{a.} Optical microscope image of an MZI with 1-mm-long electrodes. \textbf{b.} SEM image of a directional coupler. \textbf{c.} Optical microscope image of a ring resonator with a radius of 120 µm. \textbf{d.} Normalized transmission of the ring resonator. The background variation is due to the grating coupler response.}
\label{fig3:passive_optics}
\end{figure}

We fabricate the device using a thin film of MgO-doped lithium niobate bonded to a C-cut sapphire substrate. To define the waveguides, we use electron beam lithography (JEOL JBX-6300FS, 100-keV) with hydrogen silsesquioxane (HSQ) resist. We then transfer the pattern onto the LN through argon ion mill etching, clean the chip and deposit a 750-nm-thick silica cladding through plasma-enhanced chemical vapor deposition (PECVD) and anneal at 500 °C. To form the trenches we use photolithography and inductively coupled plasma (ICP) etching, and for the electrodes we do another photolithography step to liftoff 200 nm of Al evaporated at 45° for good sidewall coverage. Fig \ref{fig3:passive_optics}.a shows an optical microscope image of an MZI with 1-mm long EO modulation region (We report our results on a 3-mm device. A shorter device is displayed here for clarity.), and Fig \ref{fig3:passive_optics}.b shows a scanning electron microscope (SEM) image of a directional coupler, in which the waveguides taper down to $\text{w}_{\text{top}} = 520$ nm at the coupling region to reduce the coupler length to 37 µm. To choose this length, we fabricate a number of path-length mismatched MZIs and select the one that produces the highest-contrast fringes on bar-arm transmission. We use Euler bends to minimize intermodal scattering \cite{eulerSiN, eulerBends} and choose a minimum local bend radius of 120 µm across all devices to minimize radiation. We fabricate microring resonators with this radius, as shown in Fig. \ref{fig3:passive_optics}.c, to bound the losses through the quality factor. We couple TE polarized light into the resonator to make transmission measurements as shown in Fig. \ref{fig3:passive_optics}.d, and at $\lambda = 780$ nm we find an intrinsic quality factor of $4.4\times10^5$, which corresponds to a linear propagation loss of 1.6 dB/cm.

\section{Electro-optic Mach-Zehnder Interferometer}
\begin{figure}[!ht]
\centering\includegraphics[width=10cm]{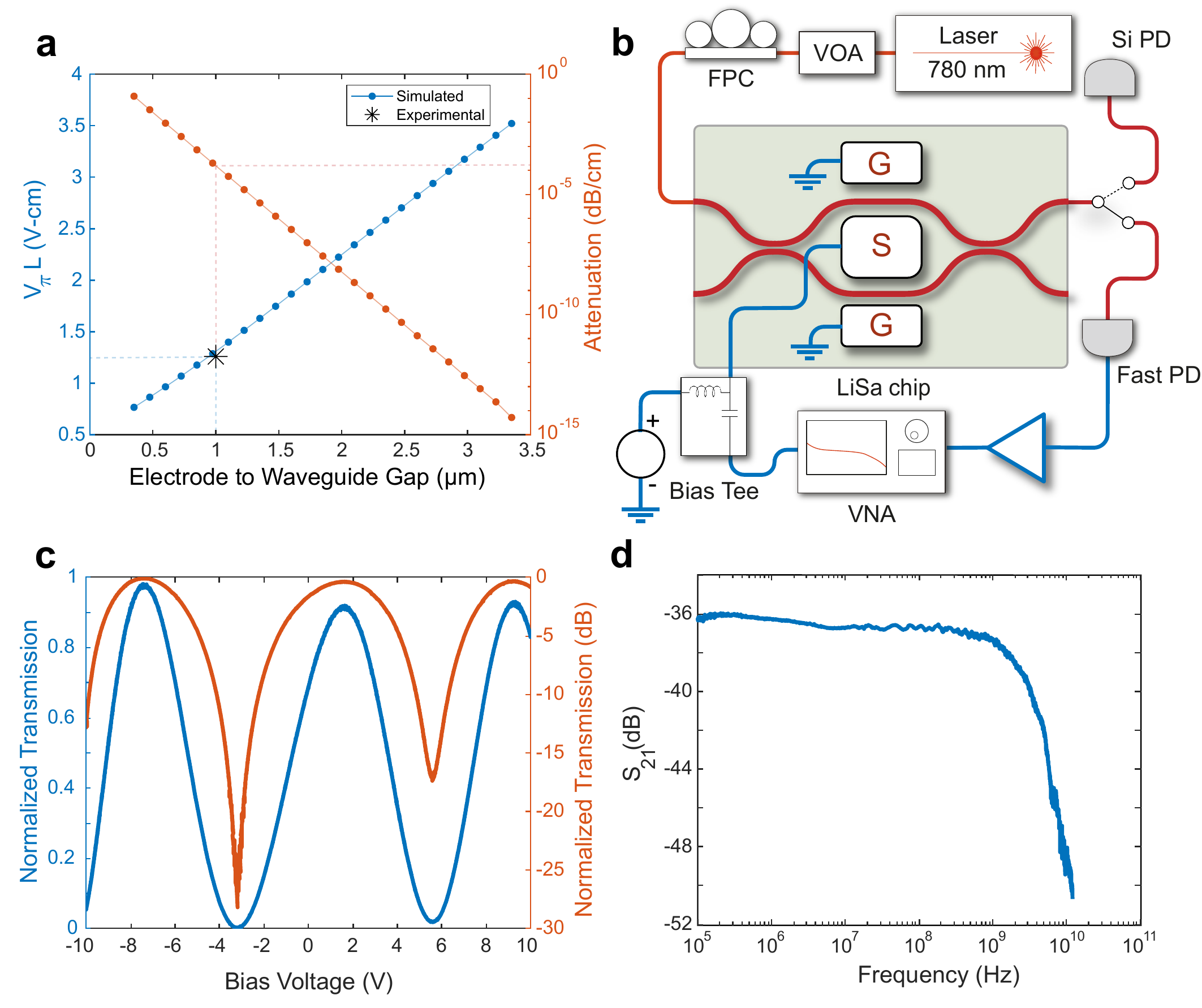}
\caption{\textbf{a.} Electrode-to-waveguide gap sweep simulation for the selection of electrode spacing. $V_{\pi}L$ increases linearly with the gap since the dominant z-component of the electric field $\mathcal{E}_{EO, z}$ is inversely proportional to the gap. The attenuation decreases exponentially as the Al is brought away from the waveguide following the roughly exponential decay of modal power with distance to the waveguide. \textbf{b.} Schematic of the measurement setup. The 780 nm laser is attenuated using a Variable Optical Attenuator (VOA) and TE-polarized using a Fiber Polarization Controller (FPC). A GSG microwave probe delivers the modulation signal from the VNA to the LiSa chip, biased at the maximum rate of change of transmission for $S_{21}$ measurements using a bias tee. The modulated light is measured by a fast photodiode with a bandwidth of 14 GHz, and the signal is amplified by a microwave amplifier. \mbox{\textbf{c.} Normalized} transmission on the bar-arm with respect to applied bias voltage showing a half-wave voltage of 4.2 V for a 3-mm long device with a 27 dB extinction ratio. \mbox{\textbf{d.} $S_{21}$} measurements using the VNA show a 3-dB bandwidth of 2.7 GHz. }
\label{fig4:eo_results}
\end{figure}

We employ a first order perturbation theory approach to calculate the EO modulation efficiency for mode $j$, given by the effective index shift $\Delta n_{\text{eff}, j}$, in the design of the interferometer. Following a reciprocity argument given in \cite{snyder2012optical}, we find the effective index shift for mode $j$ propagating along $\hat{\mathbf{y}}$ with normalized electric and magnetic fields $(\mathbf{e}_j, \mathbf{h}_j)$ as in Eq. (\ref{Eq1}).

\begin{align}
    \Delta n_{\text{eff}, j} = 
    \frac{c}{2} \frac{\int_S dA\: \mathbf{e}_j^* \cdot \epsilon_0 \Delta \bm{\epsilon} \cdot \mathbf{e}_j}{\int_S dA\: \hat{\mathbf{y}}\cdot(\mathbf{e}_j \times \mathbf{h}_j^*) },
    \label{Eq1}
\end{align}
where $S$ is the surface of the cross-section shown in \ref{fig2:optics-design}.a, $\epsilon_0$ is the free space permittivity,  $c$ is the speed of light, and $\Delta \epsilon$ is the perturbation of the dielectric tensor due to the EO effect \cite{yariv2007photonics}, given by Eq. (\ref{Eq2})
\begin{align}
    \Delta \bm{\epsilon} =  \bm{\epsilon}^T (r \cdot \bm{\mathcal{E}_{\text{eo, 1V}}}) \bm{\epsilon},
    \label{Eq2}
\end{align}
where $r_{\text{ijk}}$ is the electro-optic tensor and $\bm{\mathcal{E}}_{\text{eo, 1V}}$ is the modulating E-field created by applying $1\text{ V}$ on the signal electrode. The phase difference of the MZI arms is related to the EO modulation efficiency by $ \Delta \phi = 2\Delta n_{\text{eff, TE}_0} k_0 VL$, where $k_0$ is the free space wavevector, $V$ is the voltage applied to the signal electrode and $L$ is the modulator length, so the voltage-length product is given by $V_{\pi}L = \pi/(2 \Delta n_{\text{eff, TE}_0} k_0)$. To choose the electrode spacing, we calculate the $V_{\pi}L$ along with the attenuation due to metal absorption with a sweep over the gap between the electrode and the waveguide, as given in Fig. \ref{fig4:eo_results}.a. We select the waveguide-to-electrode gap as 1 µm, where the attenuation is simulated to be around $10^{-4}$ dB/cm, to minimize $V_{\pi}L$ without incurring optical attenuation. 

Our optical transmission measurement setup consists of a TLB6712 laser, a Variable Optical Attenuator (VOA), a Fiber Polarization Controller (FPC) and a Newport 1623 photodetector. We characterize the RF performance of the MZI through $S_{21}$ measurements with a R\&S ZNB20 Vector Network Analyzer (VNA) and a fast photodiode (Thorlabs DX12CF) with 12 GHz bandwidth followed by a microwave amplifier as shown in Fig. \ref{fig4:eo_results}.b. The device has 1 mW of optical power travelling in the waveguides during measurements, and the applied RF power is set to 8 dBm.

At $\lambda = 780$ nm, for a 3-mm long device we measure a CMOS voltage-level compatible half-wave voltage of $V_{\pi} = 4.2$ V, which corresponds to a voltage-length product of 1.26 V-cm, as shown in Fig. \ref{fig4:eo_results}.c. We observe transmission drift after changing the bias point over few seconds, which we expect is due to charge accumulation in oxide trap sites in the cladding. Once the bias has settled, we observe no drift in the modulator behavior, with or without an applied AC signal (>10 Hz). The extinction ratio (ER) is 27 dB, but we observe a reduction at higher voltages. The decrease of peak transmission along with the increase in the minimum in Fig. \ref{fig4:eo_results}.c suggests an increased loss in one of the MZI arms. ER is ideally limited by imbalance in directional coupler splitting ratio and can be increased by orders of magnitude using cascaded MZIs \cite{Liu:17} up to a limit imposed by laser phase noise \cite{West:20}.

At higher modulation frequencies, the electro-optic response is characterized by measuring the voltage modulation on a photodiode detecting the light emitted from the device with a modulated input voltage. The ratio between these voltages is called $S_{21}$. The results of the $S_{21}$ measurements are given in Fig. \ref{fig4:eo_results}.d. The EO bandwidth of the 3-mm device is found to be 2.7 GHz, which can be improved through adjusting the electrode design, increasing the electrode gap to reduce the parallel plate capacitance of the device at the expense of increased $V_{\pi}L$, and optical-microwave group velocity matching \cite{groupIndex}. Assuming an RC circuit model for the modulator, we confirm from simulations that the 3-dB roll-off should in fact occur at 2.7 GHz due to the signal-to-ground capacitance, the 50 $\Omega$ input resistance and $\approx$ 20 $\Omega$ electrode resistance due to aluminum sheet resistivity \cite{Al_resistivity}.  As expected, we observe higher EO bandwidths on shorter devices with increased $V_{\pi}$.

\section{Conclusion \& Outlook}
We demonstrated a low loss, high bandwidth, CMOS voltage-level compatible Mach-Zehnder interferometer for visible near-IR light on lithium niobate on sapphire. Our microring resonators have a quality factor of $Q = 4.4\times 10^5$, which translates to a linear propagation loss of 1.6 dB/cm. The EO bandwidth of our 3-mm-long MZI is 2.7 GHz, limited by oxide capacitance that can be reduced by distancing the electrodes (at the cost of higher $V_{\pi}L$). Adjustments to the electrode design can further improve the EO bandwidth. The voltage-length product of $V_{\pi}L = 1.26$ V-cm of the device allows small device footprints (0.35 mm$^2$) while retaining CMOS voltage-level compatibility. The extinction ratio of 27 dB at negative bias voltages and mW-level power handling ensure that the MZI can be cascaded with high output contrast. The scalable nature of our device makes the integration of deeper photonic circuits with AMO quantum computing platforms possible.

\section*{Funding}
Defense Advanced Research Projects Agency (LUMOS Program); U.S. Department of Energy (DE-AC02-76SF00515); National Science Foundation (DGE-1656518, ECCS-2026822).

\section*{Acknowledgment}
The authors would like to thank Vahid Ansari for assistance in the optical measurement setup, Alexander Y. Hwang for assistance in the acid cleaning process and Taewon Park for assistance in the fabrication of through-oxide vias.

The authors wish to thank NTT Research for their financial and technical support. Part of this work was performed at the Stanford Nano Shared Facilities (SNSF), and Stanford Nanofabrication Facility (SNF). SNSF is supported by the National Science Foundation under award ECCS-2026822. J.F.H. acknowledges support from the National Science Foundation Graduate Research Fellowship Program (Grant No. DGE-1656518).

\section*{Disclosures}
The authors declare no conflicts of interest.

\section*{Data Availability}
Data underlying the results presented in this paper are not publicly available at this time but may be obtained from the authors upon reasonable request.

\bibliography{bibliography}






\end{document}